\documentclass[prd,aps,preprintnumbers,nofootinbib,superscriptaddress,floatfix]{revtex4} 
\usepackage{epstopdf}

\usepackage{amsmath}
\usepackage{amssymb}
\usepackage{graphicx,epsfig,wrapfig,bbm}
\usepackage{bm}
\usepackage{placeins}
\usepackage{float}
\usepackage[normalem]{ulem} 
\usepackage{color}

\renewcommand\sout{\bgroup \color{blue} \ULdepth=-.5ex \ULset}

\newcommand{\bT}{b_{\bot}}
\newcommand{\kT}{k_{\bot}}

\begin{document}

\title{The transverse structure of the pion in momentum space inspired by the AdS/QCD correspondence}

\newcommand*{\Upavia}{Dipartimento di Fisica, Universit\`a degli Studi di Pavia, I-27100 Pavia, Italy}\affiliation{\Upavia}
\newcommand*{\INFN}{Istituto Nazionale di Fisica Nucleare, Sezione di Pavia,  I-27100 Pavia, Italy}\affiliation{\INFN}
\newcommand*{\Nikhef}{Nikhef, Science Park 105, NL-1098 XG Amsterdam, The Netherlands}\affiliation{\Nikhef}
\newcommand*{\Uamsterdam}{Department of Physics and Astronomy, VU University Amsterdam, De Boelelaan 1081, NL-1081 HV Amsterdam, The Netherlands}\affiliation{\Uamsterdam}

\preprint{Nikhef 2017-015}

\author{Alessandro Bacchetta}\email{alessandro.bacchetta@unipv.it}\affiliation{\Upavia}\affiliation{\INFN}

\author{Sabrina Cotogno}\email{scotogno@nikhef.nl}\affiliation{\Nikhef}\affiliation{\Uamsterdam}

\author{Barbara Pasquini}\email{barbara.pasquini@unipv.it}\affiliation{\Upavia}\affiliation{\INFN}

\begin{abstract}
We study the internal structure of the pion using a model inspired by the
AdS/QCD correspondence. 
The holographic approach provides  the light-front wave function (LFWF) for the leading Fock state component of the pion.
We adopt two different forms for the LFWF derived from the AdS/QCD soft-wall model, with free parameters
 fitted to  the available experimental information on the pion electromagnetic form factor and the leading-twist parton distribution function. 
The intrinsic scale of the model is taken as an additional fit parameter.
 Within this framework, we provide predictions for the unpolarized 
transverse momentum dependent parton distribution (TMD), and discuss its property both at the scale of the model and after TMD evolution to higher
 scales that are relevant for upcoming experimental measurements.
\end{abstract}
\maketitle

\section{Introduction}
\label{intro}
Light-front holographic Quantum ChromoDynamics
(QCD)~\citep{Brodsky:2006uqa,deTeramond:2008ht} is based on the
connection between strongly-coupled QCD in standard Minkowski space-time 
and a weakly interacting theory of gravity in higher dimensional
Anti-de Sitter (AdS) space-time. This connection is
usually referred to as AdS/QCD correspondence and is inspired by the analogous
AdS/CFT correspondence~\citep{Maldacena:1997re}, where CFT stands for
Conformal Field Theory. In the so-called ``soft-wall'' version of AdS/QCD
correspondence~\citep{Karch:2006pv}, conformal invariance is broken by
introducing 
a harmonic confining potential (whose strength is determined by a mass
parameter $\kappa$), corresponding to an infrared distortion of the AdS space.
 
Light-front holographic QCD methods 
(see \citep{Brodsky:2014yha} and references therein for a complete 
review on the topic) 
have  been employed in a number of recent works to 
obtain new insights into the structure of
hadrons~\citep{Erlich:2005qh,DaRold:2005mxj,Brodsky:2006uqa,deTeramond:2008ht,Karch:2006pv,Brodsky:2014yha}. 
One of the remarkable achievements of light-front holographic QCD 
has been to provide expressions for the light-front wave
function (LFWF) of the valence Fock-state component of mesons. 
This makes it possible to obtain direct information about many hadronic
observables, 
which can be expressed in terms of overlaps of LFWFs. 

The expressions of the LFWFs coming from the soft-wall model of the AdS/QCD
correspondence were originally derived in two different matching
procedures~\citep{Brodsky:2007hb,Brodsky:2011xx}. These two forms for the LFWF
have been used as starting point to calculate collinear and
transverse-momentum dependent parton distributions (PDFs
and TMDs, respectively), generalized parton distributions  (GPDs) and other parton
densities both for mesons and nucleons (see for instance~\citep{Forshaw:2012im,Vega:2009zb,Gutsche:2014zua,Swarnkar:2015osa,Ahmady:2016ufq,Chakrabarti:2013gra,Gutsche:2013zia,Mondal:2015uha,Chakrabarti:2015ama,Liu:2015jna,Aghasyan:2014zma,Maji:2015vsa,Maji:2016yqo,Chakrabarti:2016yuw}). 

The  structure of the pion has attracted interest since  the pion was predicted and detected experimentally.
The most intriguing aspect is the dual nature of the pion~\citep{Horn:2016rip}.
 It can be seen  as the simplest realization of a QCD bound state of quark and anti-quark as well
as the Nambu-Goldstone boson of the dynamically broken Chiral Symmetry in QCD.  
These complementary pictures  emerge when we study different properties of  the pion's interior, such as 
elastic and transition electromagnetic form factors (see
e.g.~\citep{deMelo:2003uk,deMelo:2005cy,Chang:2013nia,Arriola:2010aq,Dorokhov:2013xpa}),
distribution amplitude (see e.g.~\citep{Radyushkin:2009zg,Dumm:2013zoa}), PDFs
(see e.g.~\citep{Chang:2014lva,Chouika:2016cmv,Chen:2016sno}), GPDs (see
e.g.~\citep{Mukherjee:2002gb,Tiburzi:2002tq,Ji:2006ea,Frederico:2009fk,Dorokhov:2011ew,Fanelli:2016aqc,Mezrag:2014jka}),
TMDs (see e.g.~\citep{Pasquini:2014ppa,Lorce:2016ugb,Noguera:2015iia}), and
Fragmentation Functions (see,
e.g.,~\citep{Bacchetta:2002tk,Bacchetta:2007wc,Matevosyan:2011vj,Nam:2011hg}). 
The comparison with experiment is crucial to draw definitive conclusions, and the experiments planned at JLab 12~\citep{Dudek:2012vr},
and the  new mesonic Drell-Yan measurements
at modern facilities~\citep{Holt:2000cv,Gautheron:2010wva} 
can provide valuable information.
             
  In this work, we use the LFWFs from the AdS/QCD correspondence to study the 3D internal structure and
dynamics of the pion in momentum space. 
At leading twist, the  pion transverse momentum dependent quark-quark correlator consists of two functions, the unpolarized TMD function
$f_1(x,\mathbf{k}^2_\perp)$ and the Boer-Mulders TMD function
$h_{1}^\perp(x,\mathbf{k}^2_\perp)$. We restrict ourselves to discuss the
unpolarized TMD, since the Boer-Mulders function would require to construct a
spin-dependent LFWF, which is not naturally present in the
original AdS/QCD approach (see for example the phenomenological pion LFWF of
\citep{Gutsche:2014zua} and \citep{Ahmady:2016ufq}). 
 
A crucial ingredient of the calculation is to identify the energy scale where the
model is valid.
Light-front holographic QCD  describes 
the nonperturbative regime of QCD and therefore is
expected to be valid at low energies, approximately of the order of hadron
masses. As soon as the energy increases, a gradual transition to the
regime of perturbative QCD takes place (see
\citep{Erlich:2005qh,Deur:2005cf,Brodsky:2010ur,Deur:2014qfa,Deur:2016tte} for
more details on the transitions from one description to the other).  
In our work, we assume that the transition between the nonperturbative regime
(where the model is applicable) and the perturbative regime (where pQCD is
applicable) occurs at a precise scale, which we define as the model scale.
We fix this scale by fitting the pion PDF to available phenomenological
parametrizations, after applying DGLAP evolution
equations~\citep{Altarelli:1977zs,Dokshitzer:1977sg}. 
Once the initial scale of the model is fixed, we also discuss the application
of pQCD-based TMD evolution equations~\citep{Collins:2011zzd} 
to the unpolarized pion TMD. 

The outline of this work is as follows: in Section \ref{pionLF} we 
introduce the explicit expressions for two forms of the LFWFs in the AdS/QCD soft-wall model~\citep{Brodsky:2007hb,Brodsky:2011xx}.
After introducing the quark mass in a Lorentz invariant way  and deriving analytical expressions for the relevant hadronic matrix elements, we fix the free parameters 
of the LFWFs and the scale of the model by fitting simultaneously the data of the form factor \citep{Amendolia:1986wj,Brauel:1979zk,Volmer:2000ek,Bebek:1977pe} and phenomenological parametrizations of  the pion
PDF \citep{Wijesooriya:2005ir}. 
With these sets of parameters, we provide predictions for the unpolarized TMD at the model scale
in  Section \ref{sec:tmd}.
In Section \ref{sec:tmdev} we discuss TMD evolution~\citep{Collins:2011zzd}. 
We estimate the effects of the evolution for the broadening of the width  and the  change in the shape of the distribution, providing predictions to be tested
with upcoming experimental data from COMPASS~\citep{Gautheron:2010wva}.
In Section~\ref{conclusions} we draw our conclusions.

\section{The pion LFWF}
\label{pionLF}
Thanks to the AdS/QCD correspondence it is possible to relate the gravitational theory defined in the five-dimensional AdS space to the Hamiltonian formulation of QCD on the light-front. 
This  allows one to obtain a suitable first approximation of the valence wave function for mesons.
In particular, the direct comparison of the expression for the form factors
derived in both formalism offers the possibility of  identifying 
the spinless string modes in five-dimensional AdS space with the meson LFWFs.

In Ref.~\citep{Brodsky:2007hb},  inspired by \citep{Polchinski:2001tt,Polchinski:2002jw}, the correspondence is performed by using  the  expression  for the transition matrix element  of the free electromagnetic  current propagating in the AdS space, evaluated 
 between five-dimensional AdS modes that correspond to the incoming and outgoing hadron states in a soft-wall model effective potential.
 Taking into account only the two-parton valence  component, the explicit expression for the pion  LFWF   reads
\begin{equation}
\psi^V_{q\overline{q}/\pi}\left(x,\mbox{ }\bm{k}_{\bot}\right)\sim\frac{1}{\kappa\sqrt{\left(1-x\right)x}}e^{-\frac{1}{2}\frac{\bm{k}_{\bot}^{2}}{\kappa^{2}x\left(1-x\right)}},\label{Wave}
\end{equation}
where the superscript  $V$ indicates that we are considering the LFWF for the ``pure-valence" state of the pion.

The quark masses in the pion LFWF are included following the prescription suggested in \citep{Brodsky:2008pg}, i.e. by completing the invariant mass of the system as 
\begin{equation}
M^{2}=\sum_{i}\frac{m_{i}^{2}+\bm{k}_{\bot i}^{2}}{x_{i}}=\frac{m^2+\bm{k}^{2}_{\perp}}{x(1-x)},
\end{equation}
where $m=m_1=m_2$ and, from momentum conservation, $\bm{k}_\perp=\bm{k}_{\bot 1}=-\bm{k}_{\bot 2}$ and $x=x_1=1-x_2$.
As a result, the expression (\ref{Wave}) becomes
\begin{equation}
\psi^V_{q\overline{q}/\pi}\left(x,\mbox{ }\bm{k}_{\bot}\right)=A\frac{4\pi}{\kappa\sqrt{\left(1-x\right)x}}e^{-\frac{1}{2\kappa^{2}}\left(\frac{m^{2}}{x(1-x)}+\frac{\bm{k}_{\bot}^{2}}{x(1-x)}\right)},\label{Pion}
\end{equation}
where A is a normalization constant fixed by the condition
\begin{equation}
\label{NormWF}
\int_0^1 dx\int_{-\infty}^{+\infty} \frac{d^2\bm{k}_\bot}{16\pi^3}\,|\psi^V_{q\overline{q}/\pi}\left(x,\mbox{ }\bm{k}_{\bot}\right)|^2=1.
\end{equation}

Using the LFWF overlap representation of the PDF and form factor, we obtain
\begin{equation}
f_{1}^V(x;Q_0)=\int_{-\infty}^{+\infty} \frac{d^2\bm{k}_\bot}{16\pi^3}\,|\psi^V_{q\overline{q}/\pi}\left(x,\mbox{ }\bm{k}_{\bot}\right)|^2=A^2e^{\left({-\frac{m^{2}}{\kappa^2x}-\frac{m^{2}}{\kappa^2(1-x)}}\right)},\label{pionPDF}
\end{equation}
\begin{equation}
F_{\pi}^V(Q^2)=\int_{-\infty}^{+\infty} \frac{d^2\bm{k}_\bot}{16\pi^3}\,\psi^{*V}_{q\overline{q}/\pi}\left(x,\mbox{ }\bm{k}_{\bot}+(1-x)\bm{q}_\bot\right) \psi^V_{q\overline{q}/\pi}\left(x,\mbox{ }\bm{k}_{\bot}\right)=\int_0^1 dx A^2e^{\left({-\frac{m^{2}}{\kappa^2x}-\frac{m^{2}}{\kappa^2(1-x)}-\frac{Q^2(1-x)}{4\kappa^2x}}\right)}.\label{pionFF}
\end{equation}
where $|\bm{q}_\bot|^2=Q^2$. The condition (\ref{NormWF}) implies that $\int_0^1 dxf^V_1(x;Q_0)=F_{\pi}^V(Q^2=0)=1$. Throughout this work $f_1^q(x)=f_1^{\overline q}(x)$ is always consistently understood and we discuss results for the $\pi^+$ hadron, as the distributions for the $\pi^0$ and $\pi^-$ can be related by
 isospin and charge conjugation symmetry.

An alternative expression for the LFWF has been derived in \citep{Brodsky:2011xx}, considering the mapping of the matrix element of a confined electromagnetic current propagating in a warped AdS space to the LFWF overlap representation of the pion form factor. In this case, one obtains a LFWF which incorporates the effects due to non-valence higher-Fock states generated by the "dressed" confined current, 
and therefore  represents an ``effective" two-parton state of the pion.  
It reads 
\begin{equation}
\psi^E_{q\overline{q}/\pi}\left(x,\mbox{ }\bm{k}_{\bot}\right)\sim\, \frac{\sqrt{\log\left(\frac{1}{x}\right)}}{\kappa\left(1-x\right)}e^{ -\frac{\log(1/x)}{\left(1-x\right)^{2}}\frac{\bm{k}_{\bot}^{2}}{2\kappa^{2}}}, \label{LFWFeff}
\end{equation}
where the superscript $E$ indicates that we are considering an ``effective-valence"  component of the LFWF.
At variance with the pure-valence LFWF, the effective-valence LFWF is not symmetric in the longitudinal variables $x$ and $1-x$ of the active and spectator quark, respectively.
Introducing the quark mass dependence as outlined above, the effective-valence LFWF becomes 
\begin{equation}
\psi^E_{q\overline{q}/\pi}\left(x,\mbox{ }\bm{k}_{\bot}\right)=4\pi A\frac{\sqrt{\log\left(\frac{1}{x}\right)}}{\kappa\left(1-x\right)} e^{ -\frac{\log(1/x)}{\left(1-x\right)^{2}}\frac{\bm{k}_{\bot}^{2}+m^2}{2\kappa^{2}}} .\label{pioneff}
\end{equation}
where the parameter $A$ is once more fixed by demanding the validity of (\ref{NormWF}). The  corresponding expressions for the PDF and the form factor are given by
\begin{equation}
f_{1}^E(x;Q_0)=A^2 e^{-\frac{\log(1/x)}{(1-x)^2}\frac{m^2}{\kappa^2}};\quad F_{\pi}^E(Q^2)=\int_0^1 dx A^2 e^{-\frac{\log(1/x)}{4\kappa^2}\left(Q^2+\frac{4m^2}{(1-x)^2}\right)}.
\label{EffettivaFF}
\end{equation}

\begin{table}[t]
\centering
\label{tab:1} 
\setlength{\tabcolsep}{8pt}  
\begin{tabular}
{ccccccc}
\hline\noalign{\smallskip}
{LFWF}
& 
${m}$ {(GeV)} &
 ${\kappa}$ {(GeV)} & 

 ${Q_0}$ {(GeV)} & 

 $\chi^2_{\rm{FF}}$ & 
 $\chi^2_{\rm{PDF}}$& 
$\chi^2_{\rm d.o.f.}\left(\frac{\chi^2_{\rm{FF}}+\chi^2_{\rm{PDF}}}{N-N_{\rm{par}}}\right)$  \\
 \noalign{\smallskip}\hline\\
 					&{0.005} (fixed)  &  $0.397 \pm 0.003$ & $0.500 \pm 0.003$ & 228.7 &139.6 & 3.15 \\
$\psi^{V}_{q\overline{q}/\pi}$&{0.200 (fixed)} & $0.351  \pm 0.003$ & $0.491 \pm 0.003$ & 1064.8 & 311.6 & 11.76 \\
					&$0.0500 \pm 0.00004$ & $0.371 \pm 0.002$  & $0.498  \pm 0.002$ &  213.0 & 47.5 &2.25 \\
\noalign{\smallskip}\hline\\
								&0.005 (fixed) &  $0.261 \pm 0.002$ & $0.498 \pm 0.003$ &496.9 & 139.4 & 5.44 \\
$\psi^{E}_{q\overline{q}/\pi}$&0.200 (fixed) & $0.322  \pm 0.002$ & $0.630 \pm 0.008$ & 1349.1 &167.8 &12.96\\
								&0. (fixed)  & $0.262 \pm 0.002$  & $0.498  \pm 0.003$ & 487.4 & 142.4 &  5.38\\
\noalign{\smallskip}\hline
\end{tabular}
\caption{Results from the fit  of the pure- and effective-valence LFWFs in different quark mass scenarios. The values obtained for the $\chi^2$ are displayed separately for the form factor (FF) and PDF case (fifth and sixth column respectively). In the last column $\chi^2_{\rm d.o.f.}$ indicates the sum of the FF and PDF total values divided by the total degrees of freedom (total number of points $N$ minus the number of free parameters $N_{\rm{par}}$).}
\end{table}

We fix the parameters of the LWFs~\eqref{Pion} and \eqref{EffettivaFF}   by fitting the  available experimental data for the pion electromagnetic form factor~\citep{Amendolia:1986wj,Brauel:1979zk,Volmer:2000ek,Bebek:1977pe}  and the parametrization of the pion PDF  in \citep{Wijesooriya:2005ir}. 
For the fit of the PDF, we  apply  the DGLAP evolution equations at
next-to-leading-order (NLO) to evolve the PDF from the (low) scale of the
model $Q_0$ to the scale $Q=5$ GeV of  the parametrization, using the HOPPET
code~\citep{Salam:2008qg}. We leave in the initial scale  $Q_0$ as an additional free parameter to be fitted with the data. Starting from the functional form of the parametrization  \citep{Wijesooriya:2005ir}, we select 61 equally-spaced points from $x=0.2$ to $x=0.8$ and for each of them we construct error bars 
by propagation of the errors on the individual parameters. 
Summing the PDF points and the 58 form factor points,  we perform the fit 
using in total 119 points.
In the case of the pure-valence LFWF we consider two different fitting
strategies: either we fix the quark mass to a constant value (``current quark''
mass $m=0.005$ GeV and ``constituent quark'' mass $m=0.2$ GeV) or, alternatively, we let the quark mass entering as an additional fit parameter.
For the effective-valence LFWF, we fix the quark mass to the same values as
before, but we include also the limit of massless quarks (leaving the quark
mass as a free parameter in this case leads anyway to a vanishing mass).
The results of the fit  are summarized in  Tab.~\ref{tab:1}.  In the
following we discuss the results for two sets of parameters in Tab.~\ref{tab:1} corresponding with the lowest value of the total  $\chi^2_{\rm
  d.o.f.}$ for non-vanishing quark mass. 

In Fig. 1 we show the results for the form factor of the  pure-valence (solid curve) and effective-valence (dashed curve)  LFWF.
 The corresponding results for the PDF are shown in Fig. 2(a) and 2(b), respectively.
 The solid curves  show the results at the hadronic scale, and the dashed curves are obtained  after NLO evolution to $Q=5$ GeV. The shaded band corresponds to the results from  the parametrization at $Q=5$ GeV  of Ref.~\citep{Wijesooriya:2005ir}.
 

\begin{figure}[t]
\centering
\includegraphics[scale=0.35]{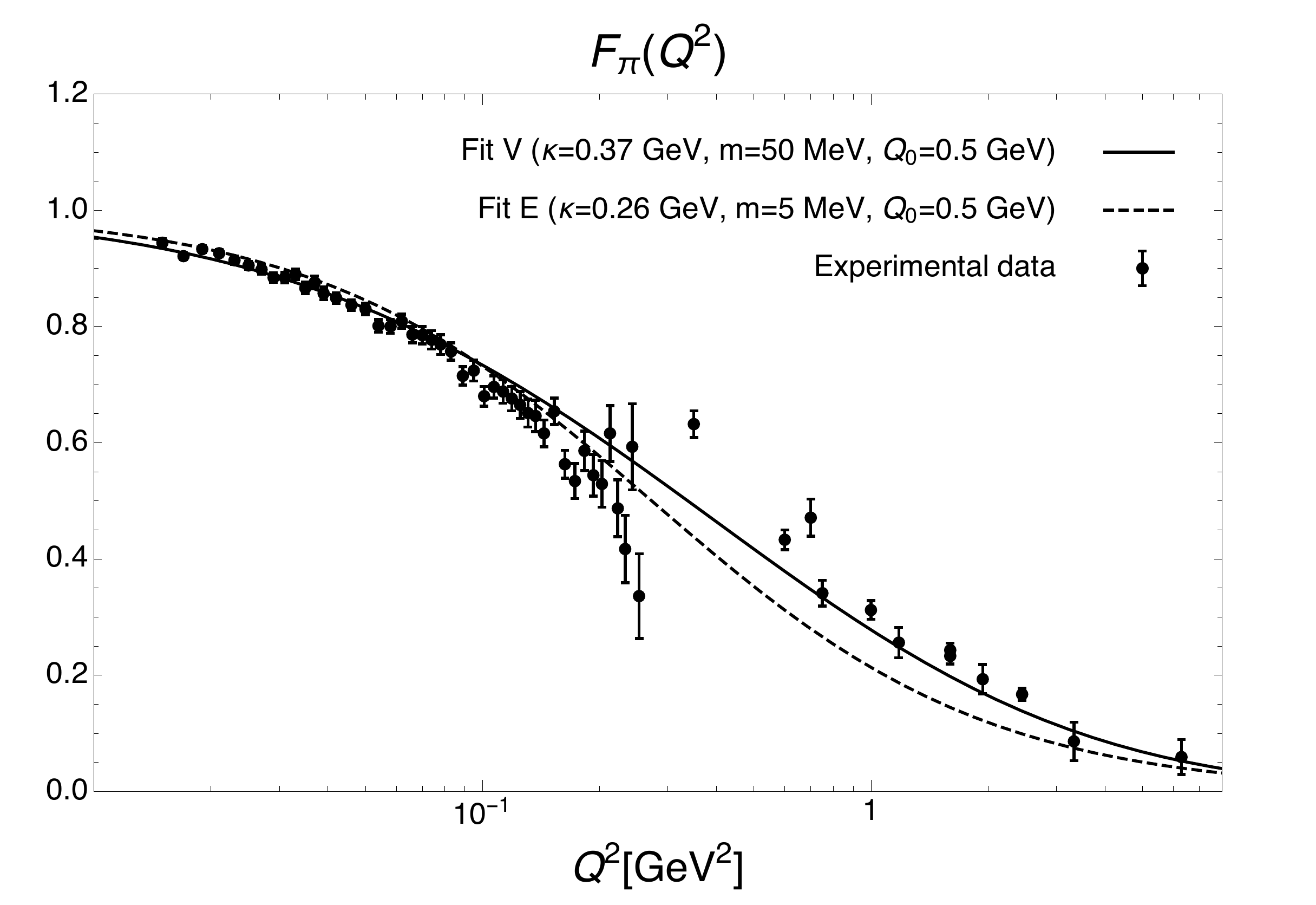}
\caption{\footnotesize{Results for the pion electromagnetic form factor from the pure-valence LFWF (solid curve) and the effective-valence LFWF (dashed curve) with the two sets of parameters in Tab.~\ref{tab:1} corresponding with the lowest values of $\chi^2_{{\rm d.o.f.}}$ for non-vanishing quark mass. The experimental data are from Refs.~\citep{Amendolia:1986wj,Brauel:1979zk,Volmer:2000ek,Bebek:1977pe}.}}
\end{figure}

\begin{figure}[h]
\begin{tabular}{ccc}
\hspace{-.2cm}\includegraphics[scale=0.35]{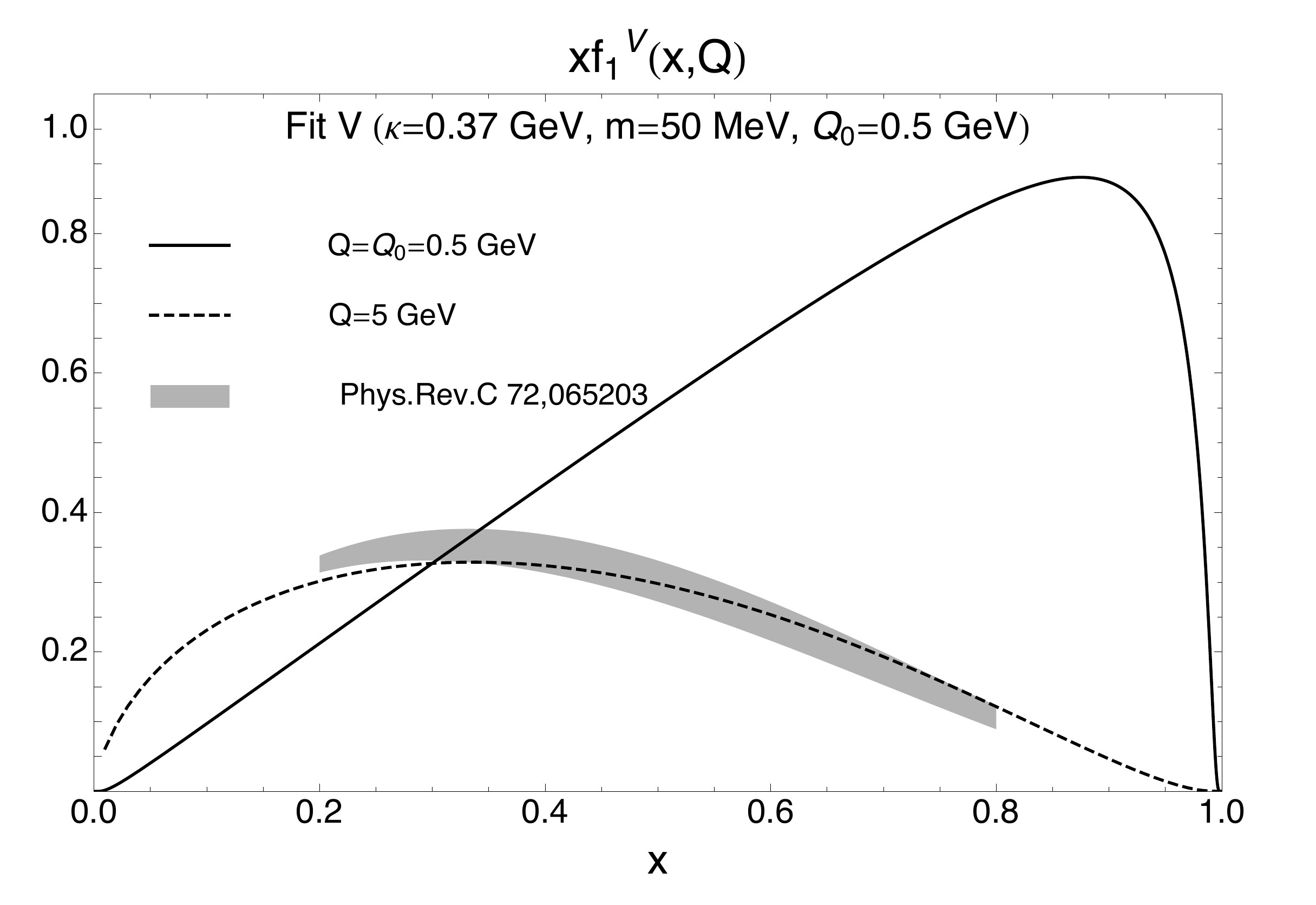}
&
&
\includegraphics[scale=0.35]{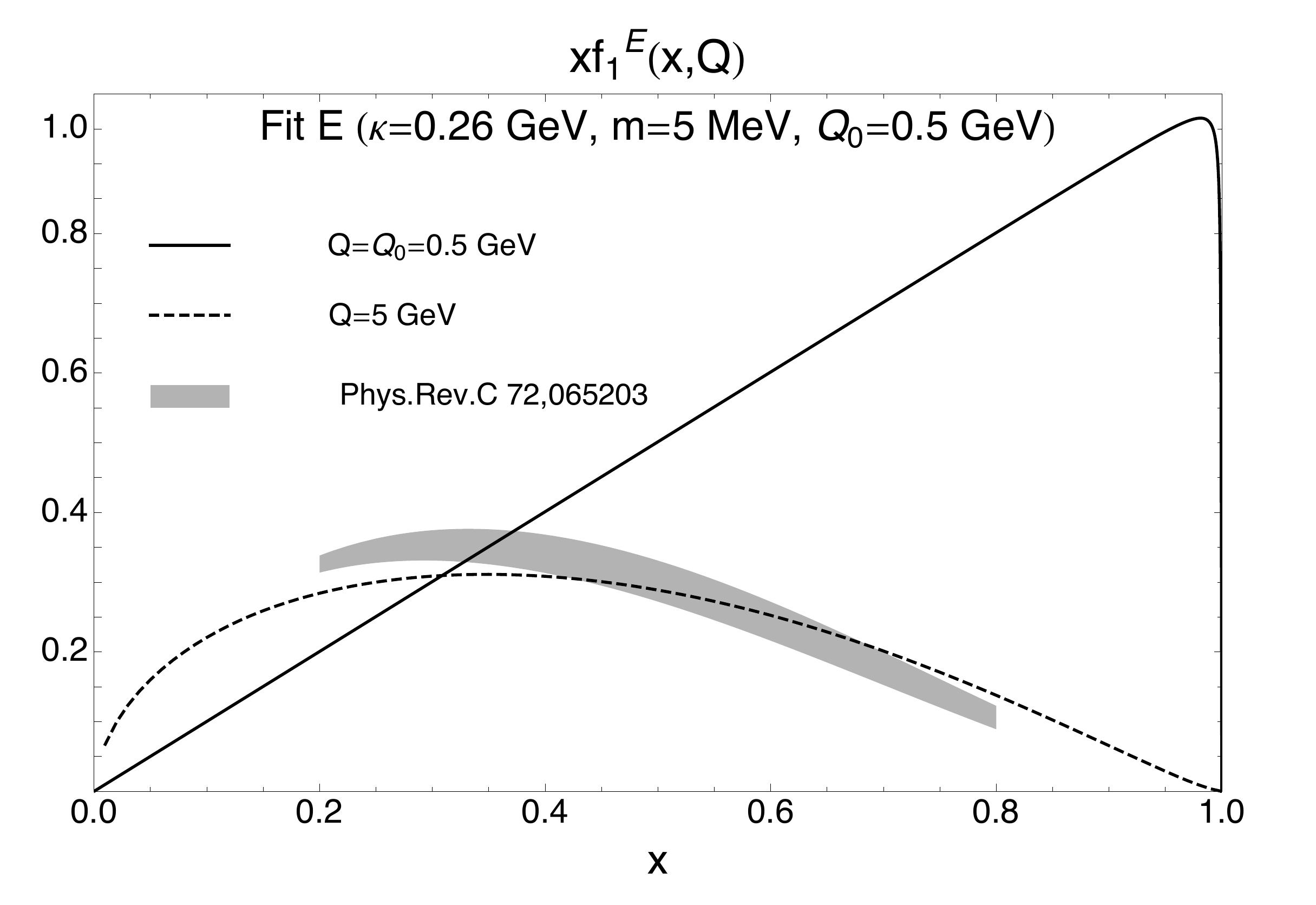}\\
\hspace{-.2cm} (a)& 
\hspace{-3.4cm}
& (b)\\
\end{tabular}
\caption{\footnotesize{Results for the quark PDF of the pion as function of $x$ from the pure-valence LFWF (a) and the effective-valence LFWF (b), with the two sets of parameters in Tab.~\ref{tab:1} corresponding with the lowest values of $\chi^2_{{\rm d.o.f.}}$ for non-vanishing quark mass.
Solid curves: results at the initial scale of the model. Dashed curves: results after NLO evolution to $Q=5$ GeV. Dashed band: parametrization at $Q=5$ GeV from Ref. \citep{Wijesooriya:2005ir}. }}
\label{fig:PDFVE}
\end{figure}

The results from the pure-valence LFWF are in good agreement with the available experimental and phenomenological information, while
 a worst comparison, especially for the form factor, is obtained in the case of the effective-valence LFWF.

The mass parameter  $\kappa$ plays a very important role, as it is originally
the only free parameter of the theory and it is related to the strength of the
confining harmonic potential in the soft-wall model
\citep{deTeramond:2008ht,Trawinski:2014msa}. The value $\kappa \approx 0.37$
GeV obtained in the pure-valence LFWF case is similar to what was obtained in
Ref.~\citep{Brodsky:2007hb},  
whereas in the study of the effective-valence LFWF we obtain smaller
values, $\kappa \approx 0.26$ GeV, compared to previous
analyses~\citep{Gutsche:2013zia,Gutsche:2014zua}.  
Moreover, we point out that a larger value, namely $\kappa=0.54$ GeV, is needed in order to describe the hadronic mass spectra and the Regge trajectories~\citep{Branz:2010ub,Colangelo:2008us,Forkel:2007cm} and this value has been quite extensively used (see \citep{Brodsky:2014yha,Ahmady:2016ufq} for a more complete overview). Recent works \citep{Brodsky:2016rvj,Deur:2016opc} quote a value of $\kappa=0.523\,\rm{GeV}$  to reproduce Regge slopes for mesons and baryons and to realize the transition from the non-perturbative (described by light-front holography) and the perturbative regimes, which occurs at an energy scale of about $1$ GeV.

Our result for the  initial scale is $Q_0\sim$ 0.5 GeV in the pure-valence case and is consistent with the values obtained in different phenomenological quark models \citep{Pasquini:2014ppa,Fanelli:2016aqc}, where the scale is fixed by requiring that the model results for the momentum carried by the valence quarks match the experimental value, after DGLAP evolution.  We also notice that  the fit of  the quark mass  provides a value that is quite close to the average effective light-quark mass obtained in LF holographic QCD from the meson spectrum \citep{Brodsky:2014yha}. 
In the case of the effective-valence LFWF,  we  expect that the inclusion of the effects of higher-order Fock state components should correspond to a higher  hadronic scale. This is the case when comparing the results 
between the effective-valence and pure-valence LFWF with $m=200$ MeV and similar values of $\kappa$.
However, for the other quark-mass scenarios we find similar values of $Q_0$ in the two models, which are compensated by much lower values for the parameter $\kappa$ in the case of the effective-valence LFWF. Both the values of $\kappa$ and the  initial scale $Q_0$ differ with respect to \citep{Brodsky:2016rvj,Deur:2016opc}.

\section{TMD analysis}
\label{sec:tmd}
\begin{figure}[h]

\centering

\begin{tabular}{ccc}
\includegraphics[scale=0.4]{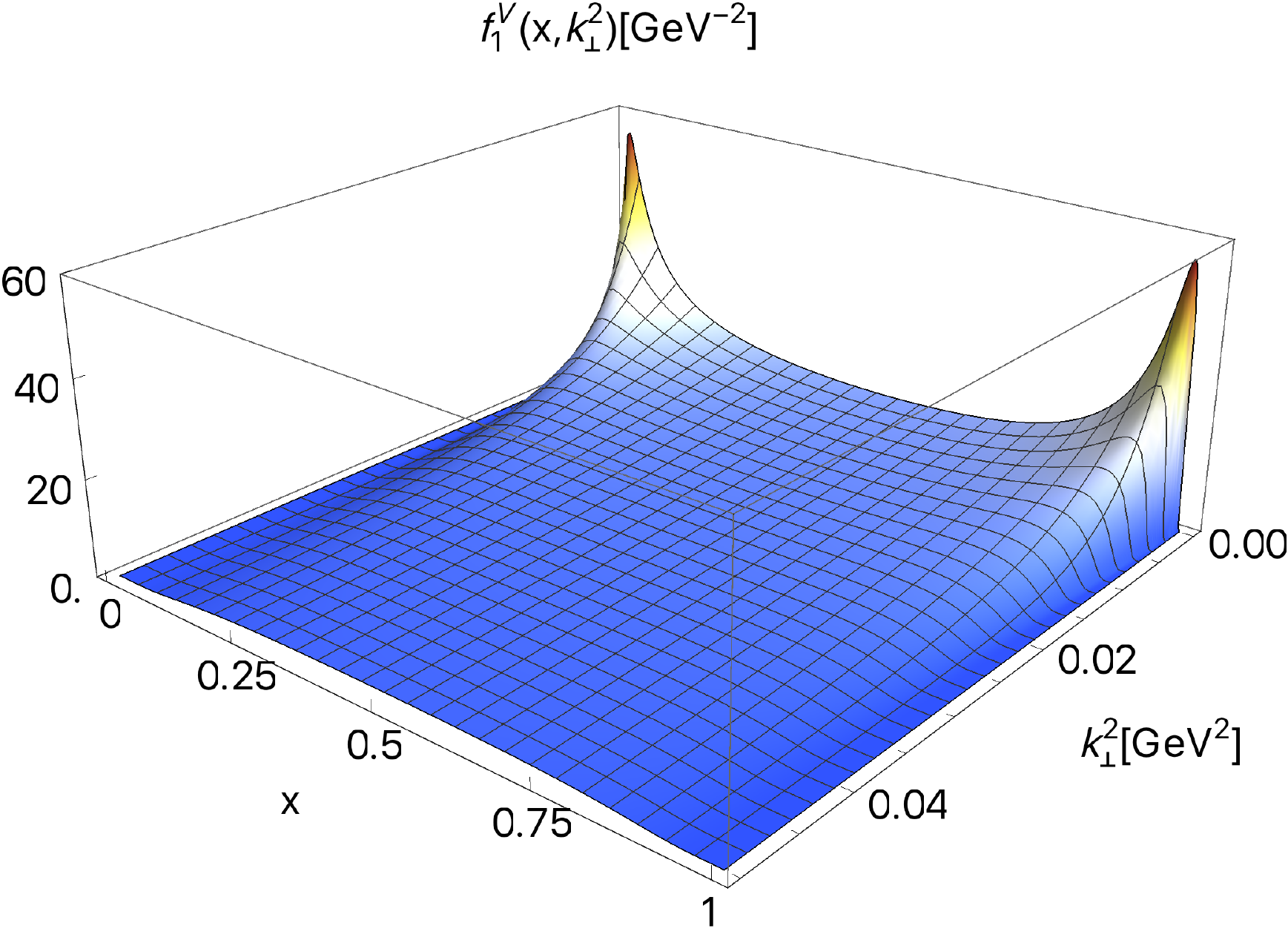}
&\quad\quad
&
\includegraphics[scale=0.4]{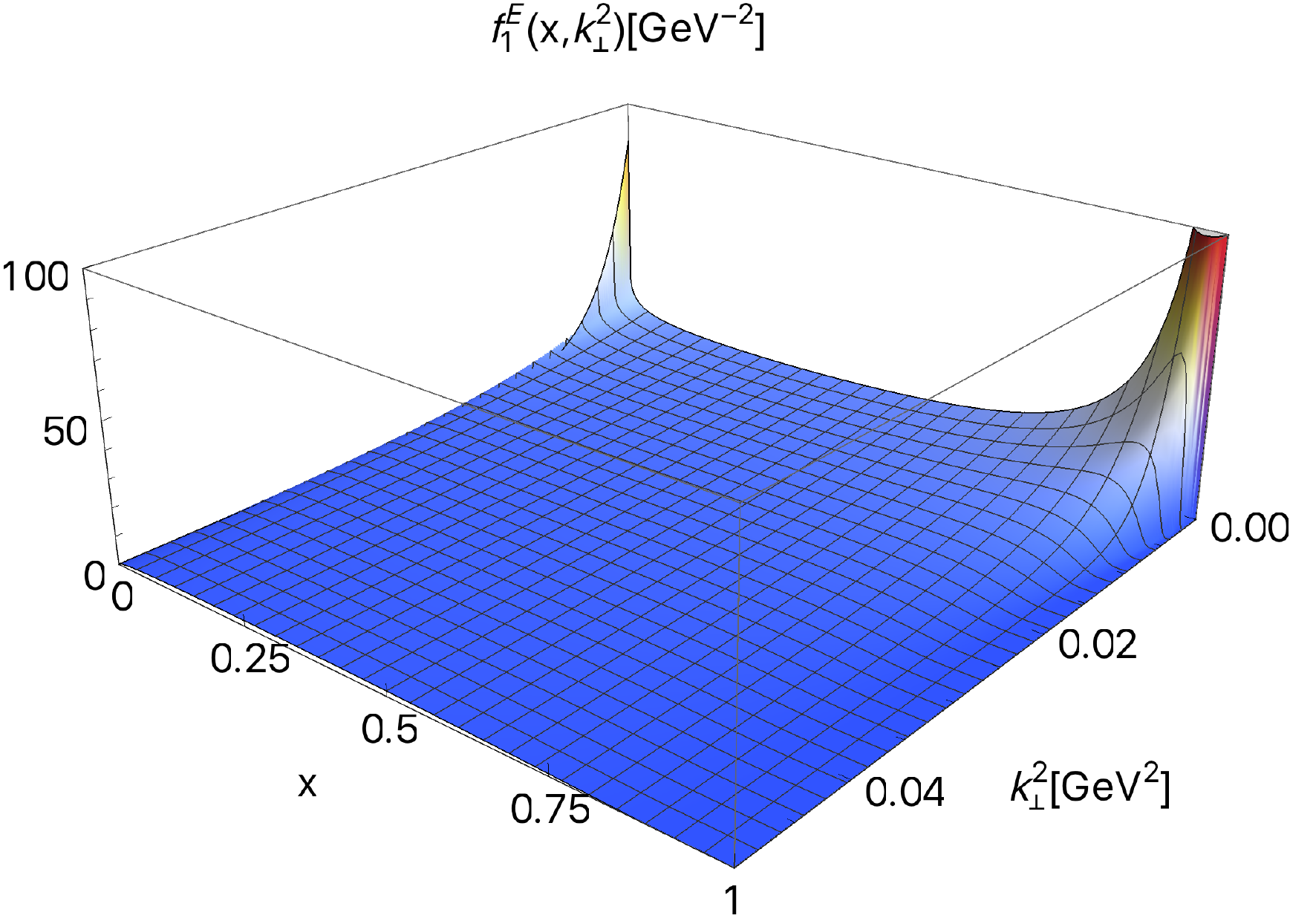}\\
 \hspace{-0.5 truecm}(a)& & \hspace{0.9 truecm} (b)\\

\end{tabular}
\caption{\footnotesize{Results for the quark TMD of the pion as function of $x$ and $k^{2}_\perp$ from the pure-valence LFWF (left) and the effective-valence LFWF (right) with the two sets of parameters in Tab.~\ref{tab:1} corresponding with the lowest values of $\chi^2_{{\rm d.o.f.}}$ for non-vanishing quark mass.}}
\label{figkTVal}

\end{figure}

The unpolarized TMD $f_1(x,\bm{k}^2_{\bot})$ can be obtained from the following LFWF overlap~\citep{Pasquini:2014ppa}
\begin{equation}
f_1(x,\bm{k}^2_{\bot};Q_0)=\frac{1}{16\pi^3}|\psi_{q\overline{q}/\pi}\left(x,\mbox{ }\bm{k}_{\bot}\right)|^2 \label{TMDboth},
\end{equation}
which reduces to the PDF in Eq.~\eqref{pionPDF} after integration over $\bm k_{\bot}$.
Using the expressions in Eqs. (\ref{Pion}) and  (\ref{pioneff}), one finds that the TMD in both models
is a Gaussian distribution  in $\bm{k}_{\bot}$, with an $x$-dependent mean square transverse momenta, i.e. 
\begin{align}
f^{V}_1(x,\bm{k}^2_{\bot};Q_0)&=\frac{ A^2 }{\pi\kappa^2 x(1-x)
}e^{-\frac{\bm{k}^2_{\bot}+m^2}{\kappa^2 x(1-x)}}, 
&
\langle k^{2}_{\bot}(x)\rangle^{V}&=\kappa^2 x(1-x),
\label{TMDValenza}
\\
f^{E}_1(x,\bm{k}^2_{\bot};Q_0)&=\frac{ A^2 \log
  \left(\frac{1}{x}\right)}{\pi\kappa^2 (1-x)^2}
e^{-\log\left(\frac{1}{x}\right)\frac{\bm{k}_{\bot}^2+m^2}{\kappa ^2
    (1-x)^2}}, 
&
\langle k^{2}_{\bot}(x)\rangle^{E}&=\frac{\kappa^2(1-x)^2}{\log (1/x)},
\label{TMDEffettiva}
\end{align}
where $k_\perp=|\bm{k}_\perp|$.
In Fig. \ref{figkTVal} we show the results for the TMD in the two models, as function of $x$ and $k^{2}_{\perp}$.
As in the case of the PDF, the pure-valence model is symmetric under the exchange of $x\rightarrow1-x$, while this symmetry  is lost when including effects beyond the valence sector in the effective-valence LFWF. 
The fall-off in $k^{2}_\perp$ is Gaussian in both models.

\begin{figure}[h]
\centering
\includegraphics[scale=0.35]{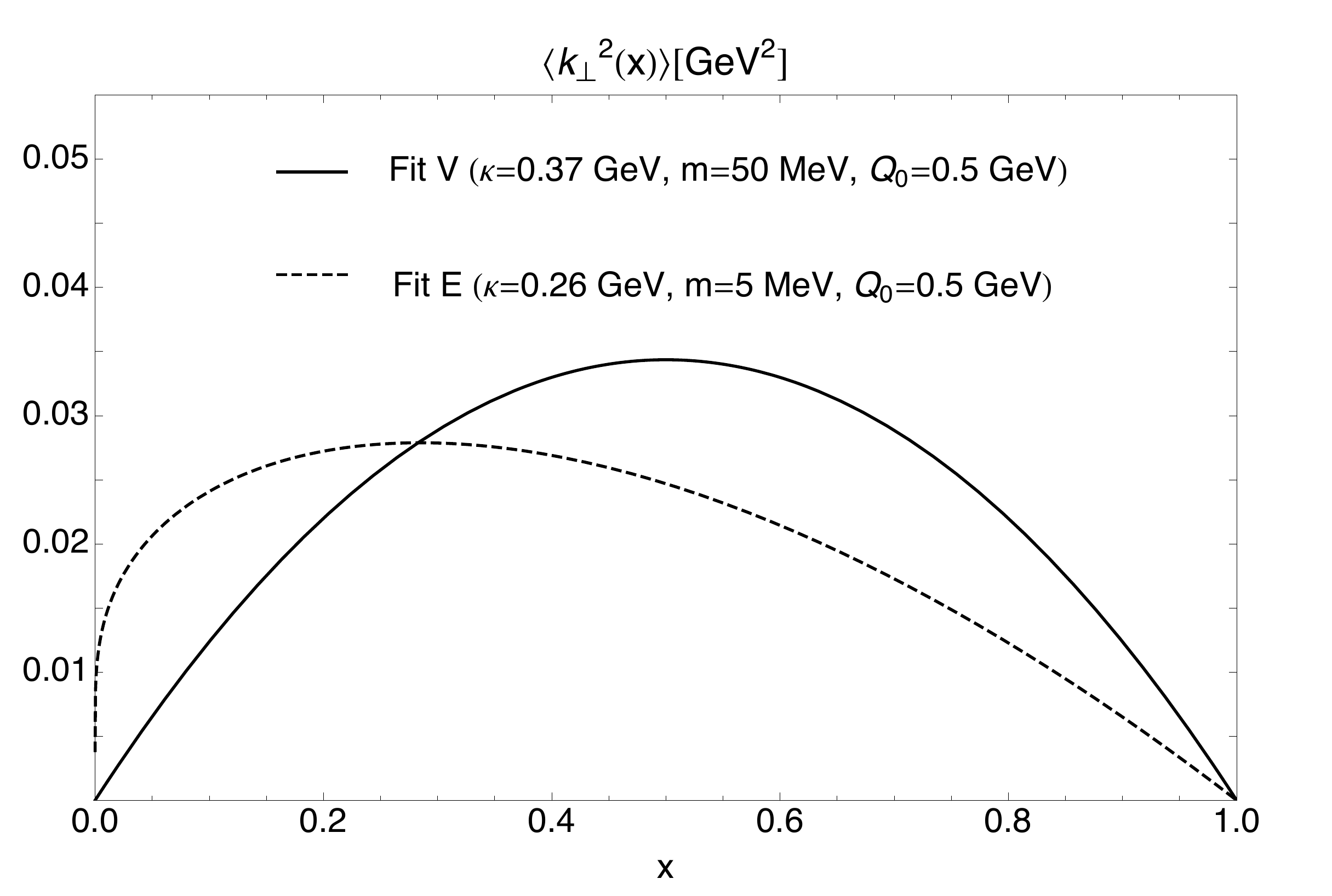}
\caption{\footnotesize{Results for the width $\langle{k}^{2}_\perp(x)\rangle$ as a function of $x$ for the pure-valence LFWF (solid curve) and the effective-valence LFWF (dashed curve),
with the two sets of parameters in Tab.~\ref{tab:1} corresponding with the lowest values of $\chi^2_{{\rm d.o.f.}}$ for non-vanishing quark mass.}}
\label{kTAvBothQ0}
\end{figure}

The width of the distribution $\langle{k}_{\bot}^2(x)\rangle$   is shown  as function of $x$  in Fig.~\ref{kTAvBothQ0}.
It is slightly larger in the pure-valence model, with a maximum at $x=0.5$ and the characteristic symmetric behaviour around the maximum.
Integrating over $x$, one obtains $\langle{k}^{2}_{\bot}\rangle^{V}=0.023\,\textrm{GeV}^2$.
In the case of the effective-valence LFWF the maximum is shifted at lower values of $x$, i.e.  $x=0.28$, and the result after $x$-integration 
is $\langle{k}^{2}_{\bot}\rangle^{E}=0.020\,\textrm{GeV}^2$.

\subsection{TMD evolution}
\label{sec:tmdev}

As explained before, the AdS/QCD LFWF and the resulting TMDs are obtained at a
scale of about 0.5 GeV. In order to be able to compare with data or
extractions, TMDs need to be evolved according to TMD evolution
equations (see, e.g., Ref.~\citep{Rogers:2015sqa}). 
These equations describe the broadening of the initial TMD due to
gluon radiation. 

Even though TMD evolution equations are based on
perturbative QCD calculations, their 
implementation requires the introduction of some
prescriptions to avoid extending the calculations outside their region of
validity. In general, such prescriptions have the effect of inhibiting
perturbative gluon radiation at low transverse momentum and at low $Q$, but
must be complemented with an additional component of gluon radiation, usually
referred to as nonperturbative component of TMD
evolution~\citep{Collins:2014jpa}.  
This component
cannot be predicted by perturbative QCD, but has to be extracted from
experimental measurements, taking advantage of the fact that 
it is highly universal (i.e., it is independent of the quark's flavor and
spin, the parent hadron, the type of process, and whether one considers TMD
distribution and fragmentation functions). It may be possible to use
AdS/QCD correspondence also to compute the nonperturbative components of TMD
evolution, but we leave this issue to future studies (for a recent example of a computation
of the behavior of the nonperturbative component of TMD evolution see
Ref.~\citep{Scimemi:2016ffw}) .

Several prescriptions have been proposed in the literature (see, e.g.,
Refs.~\citep{Collins:1984kg,Laenen:2000de,DAlesio:2014mrz,Collins:2014jpa,Bacchetta:2015ora}).
In principle, if
complemented with the appropriate nonperturbative components, they should lead
to compatible results for the evolved TMDs. However, there is still
considerable uncertainty on the nonperturbative components and systematic
studies of these uncertainty are still lacking. We therefore choose a specific
implementation of TMD evolution equations, which has been successfully applied
to the description of data in the range 1.2 GeV $\lesssim Q \lesssim$ 80
GeV. Details of this implementation are discussed in Ref.~\citep{Bacchetta:2017}. We
summarize here the most important points.

TMD evolution is implemented in the space Fourier-conjugate to
$\kT$. Therefore, we first define the Fourier-transformed TMDs
\begin{equation}
\widetilde{f}_1(x,\bT^2;\mu) = 
\int_0^{\infty} {d \kT}\kT J_0\big(\bT \kT \big)
f_1(x,\kT^2;\mu)
\end{equation} 
and we use the following form for the evolved TMDs in $\bT$ space
 (see Refs.~\citep{Collins:2011zzd,Aybat:2011zv}) 
\begin{equation}   
\widetilde{f}_1^a(x,\bT^2;\mu) =
\sum_{i=q,\bar q,g}\bigl( 
\tilde{C}_{a/i}\otimes
f_1^i\bigr)(x;\mu_b) 
e^{\tilde S(\bar{b}_{\ast};\mu_b,\mu)}
e^{g_K(\bT)\ln\frac{\mu}{Q_0}}
\widetilde{f}_{1}^a(x,\bT^2; Q_0),
\label{e:TMDevol1}
\end{equation}
where the label $a$ indicates the parton type.
We consider the above equation at Next-to-Leading Logarithmic (NLL)
approximation and at leading order in $\alpha_S$. In this case, the
convolution at the beginning of the evolved formula reduces simply to
\begin{equation} 
\sum_{i=q,\bar q,g}\bigl( C_{a/i} \otimes f_1^i \bigr) (x; \mu_b^2) \approx
f_1^a (x; \mu_b^2), 
\end{equation} 
and the expression for the Sudakov form factor 
$\tilde S(\bar{b}_{\ast};\mu_b,\mu)$ can be found, e.g., in
Ref.~\citep{Frixione:1998dw,Echevarria:2012pw}. 
We further use 
\begin{align}
\mu_b &= \frac{2 e^{-\gamma_E}}{\bar{b}_{\ast}},
&
g_K &= - g_2 \bT^2 / 2,
&
Q_0&= 0.5 \text{ GeV}.
\label{eq:g2}
\end{align}  
We introduced the following
variable
\begin{equation} 
\bar{b}_{\ast} \equiv b_{\rm max} \Bigg(\frac{1-e^{- \bT^4 / b_{\rm max}^4}}
         {1-e^{- \bT^4 / b_{\rm min}^4}} \Bigg)^{1/4},
\label{e:b*}\end{equation}
with 
\begin{align}
b_{\rm max} &= 2 e^{-\gamma_E}/Q_0 = 2.246 \text{  GeV}^{-1}, 
&
b_{\rm min} &= 2 e^{-\gamma_E}/Q \ .
\label{e:bminmax} 
\end{align} 
The above choice guarantee that at the initial scale $Q=Q_0$ any effect of TMD
evolution is absent. The model results are thus preserved and in particular
the relation between TMD and collinear PDF is maintained.

The value of the $g_2$ parameter should be extracted from experimental data,
keeping all other choices fixed. In a recent analysis, the parameter was found
to be $0.13 \pm 0.01$ GeV$^2$ in combination with a $b_{\rm max}$ that was half of the
value we assume here. Since $b_{\rm max}$ and $g_2$ are in general
anticorrelated, we choose for the present analysis the following three values
\begin{equation} 
g_2 = 0.09,\;  0.11,\;  0.13 \text{ GeV}^2.
\end{equation}

\begin{figure}[t]

\centering

\begin{tabular}{ccc}
\hspace{-.2cm}
\includegraphics[scale=0.35]{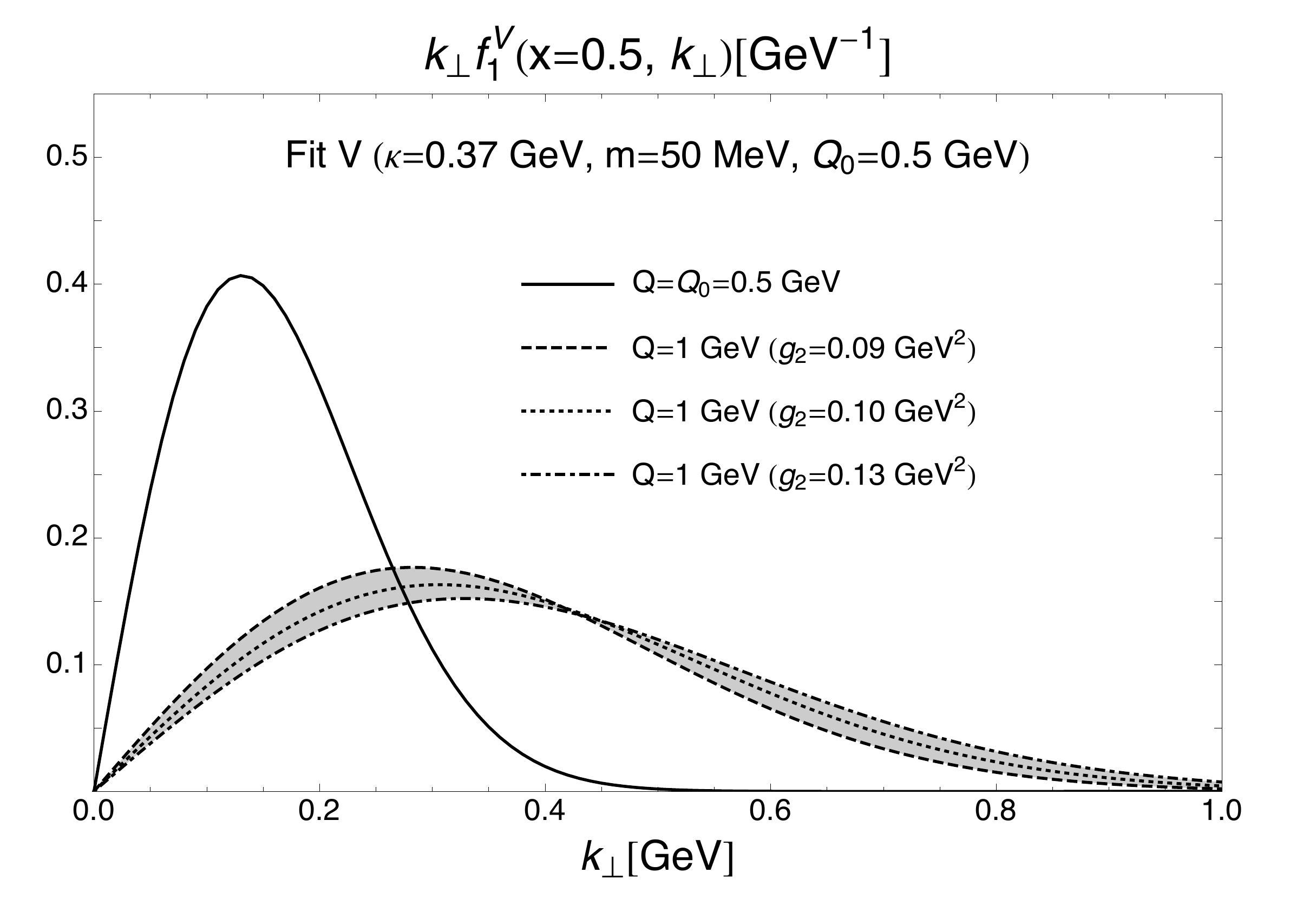}
&

&
\includegraphics[scale=0.35]{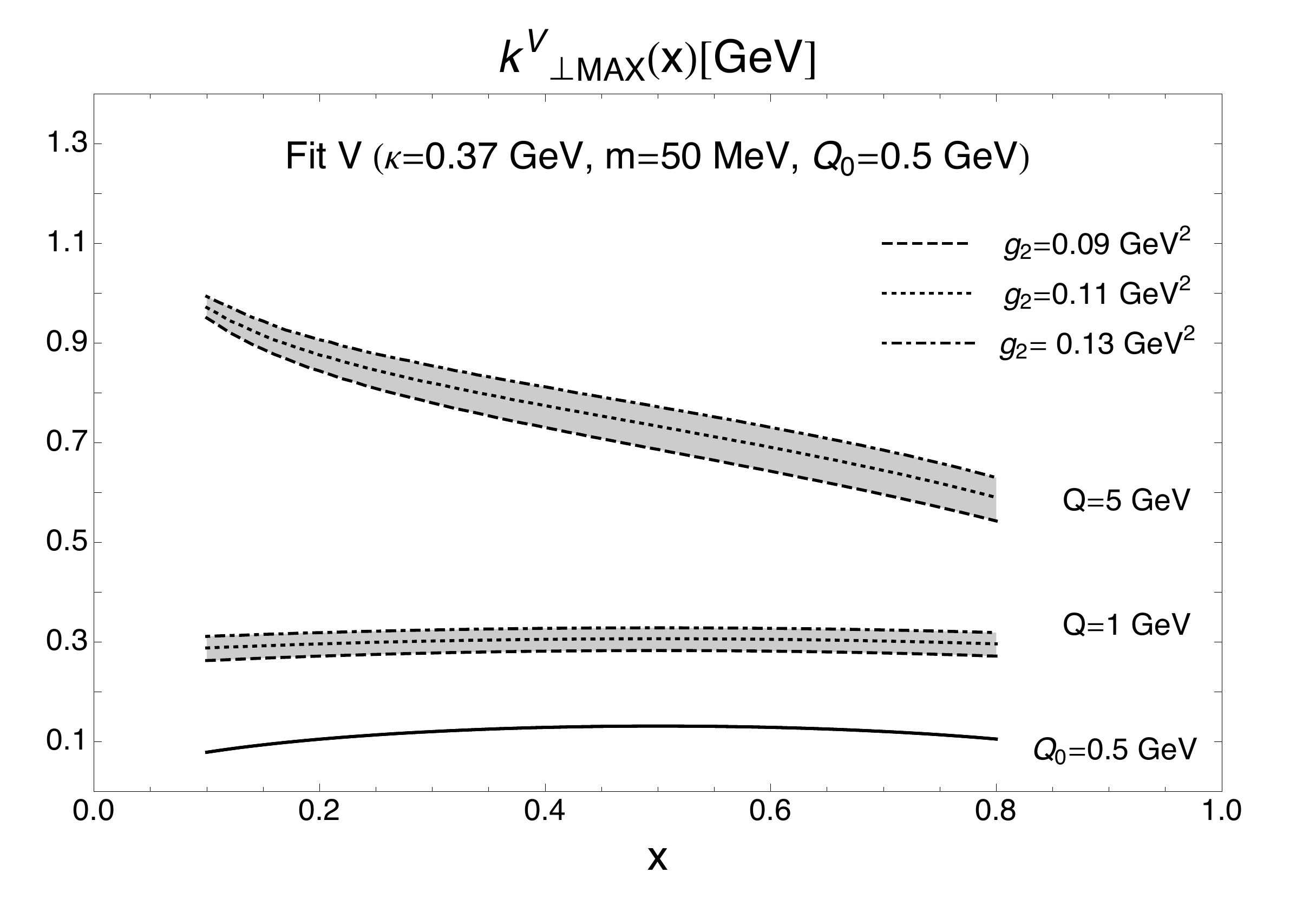}\\
\hspace{-.2cm} (a)& 
\hspace{-3.4cm}
& (b)\\
\end{tabular}
\caption{\footnotesize{Left panel: results for the quark TMD of the pion, multiplied by $k_\perp$, from the pure-valence LFWF for the $m=50$ MeV scenario, as function of $\kT$ and at fixed $x=0.5$.
The solid curve shows the result at the scale of the model, $Q_0=0.5$ GeV, corresponding with the initial scale for the TMD evolution. 
The shaded band  gives the spread of the results after evolution of the TMD to 1 GeV with three different  values of $g_2$:  0.09 GeV$^2$ (dashed curve), 0.11 GeV$^2$ (dotted curve) and 0.13 GeV$^2$ (dashed-dotted curve). 
Right panel: results for $k_{\bot MAX}$ as function of $x$, at the scale of the model (solid curve) and after 
TMD evolution to $Q=1$ GeV (lower band) and $Q=5$ GeV (upper band) with three different values of $g_2$: 0.09  GeV$^2$ (dashed curve), 0.11 GeV$^2$ (dotted curve) and 0.13  GeV$^2$ (dashed-dotted curve).  }}
\label{ValenceEvo}

\end{figure}

\begin{figure}[h]

\centering

\begin{tabular}{ccc}
\hspace{-.2cm}\includegraphics[scale=0.35]{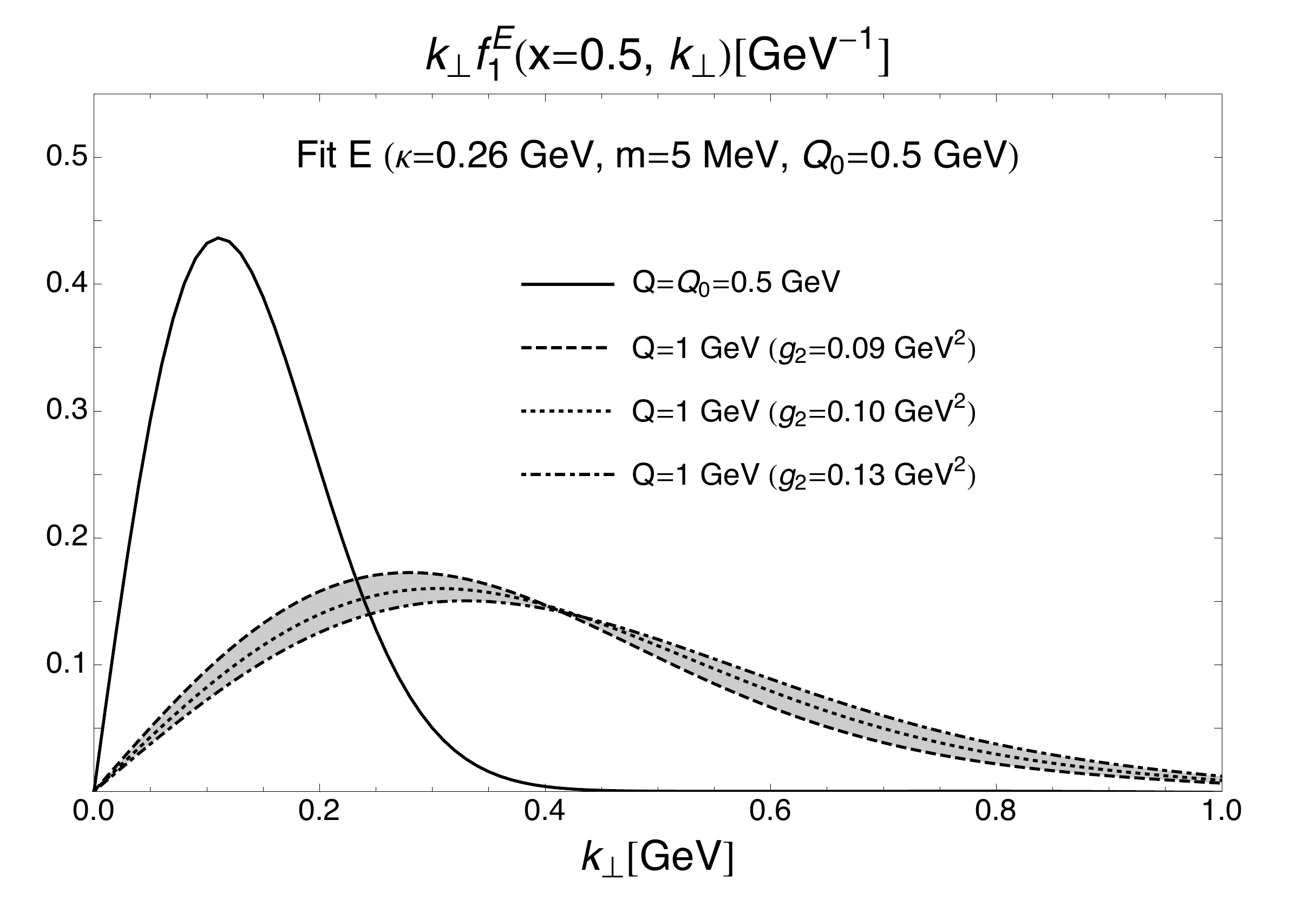}
&

&
\includegraphics[scale=0.35]{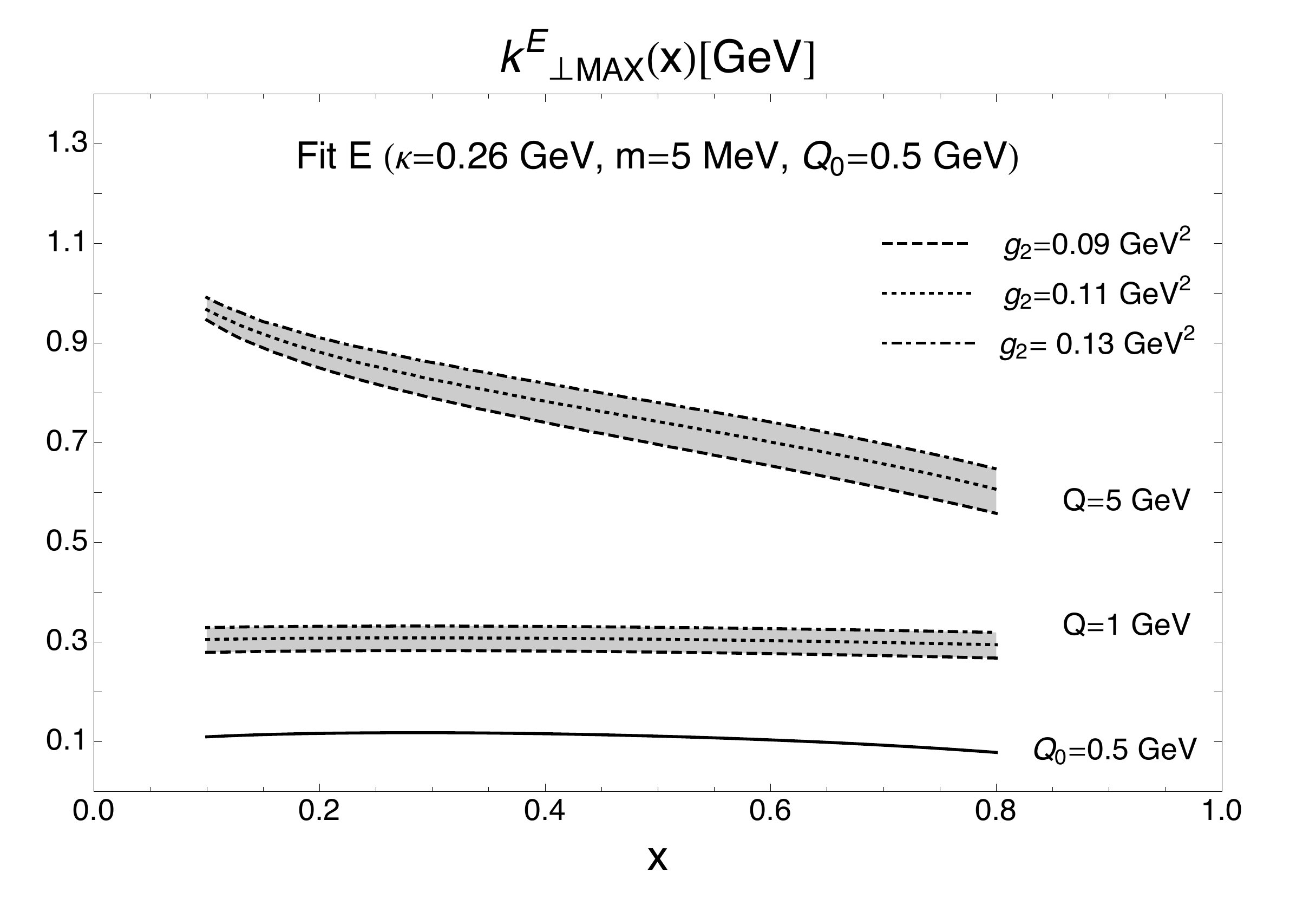}\\
\hspace{-.2cm} (a)& 
\hspace{-3.4cm}
& (b)\\

\end{tabular}
\caption{\footnotesize{
Left panel: results for  the quark TMD of the pion, multiplied by $k_\perp$, from the effective-valence LFWF for the $m=50$ MeV scenario as function of $\kT$ and at fixed $x=0.5$.
The solid curve shows the result at the scale of the model, $Q_0=0.5$ GeV, corresponding with the initial scale for the TMD evolution. 
The shaded band  gives the spread of the results after evolution of the TMD to 1 GeV with three different  values of $g_2$:  0.09 GeV$^2$(dashed curve), 0.11 GeV$^2$ (dotted curve) and 0.13 GeV$^2$ (dashed-dotted curve). 
Right panel: results for $k_{\bot MAX}$ as function of $x$, at the scale of the model (solid curve) and after 
TMD evolution to $Q=1$ GeV (lower band) and $Q=5$ GeV (upper band) with three different values of $g_2$: 0.09 GeV$^2$ (dashed curve), 0.11 GeV$^2$ (dotted curve) and 0.13 GeV$^2$ (dashed-dotted curve). } }
\label{EffectivEvo} 
\end{figure}

Figures \ref{ValenceEvo}(a) and \ref{EffectivEvo}(a) show the effect of TMD
evolution when going from the model scale to 1 GeV (at an illustrative value
of $x=0.5$).  
The value of $k_{\bot}$
corresponding to the position of the peak of the distributions
$k_{\bot}f_1(x,k_{\bot}^2)$ 
can be used as a measure of the
width of the TMDs. The peak
moves from about 0.1 to 0.3 GeV, showing 
that there is a broadening of the width of the distributions. 
Even if this not evident from the plot, the distributions are no longer
Gaussian. 

Figures
\ref{ValenceEvo}(b) and \ref{EffectivEvo}(b) show the position of the peak 
for $x$ between 0.1 and 0.8 and for three values of $Q$. At
the scale of the model, this is an analytic function which reads: 
\begin{equation}
k_{\bot MAX}(x)=\sqrt{\frac{\langle k^{2}_{\bot}(x)\rangle}{2}}.
\end{equation}
After evolution to 1 GeV, as already observed, the width of the TMD increases
to about 0.3 GeV in both versions of the model. The $x$ dependence of the TMD
width is rather flat. The symmetry about $x=0.5$ 
of the pure-valence model is lost. The two models become quite similar to each
other: the position of the peak is the same within a 5\% error.
At 5 GeV, the width of the TMD increases to about 0.7 GeV at $x=0.5$ and
increases at low $x$, and is again very similar in the two versions of the
model. 

In summary, TMD evolution from the model scale (0.5 GeV) to a typical
experimental scale of 5 GeV increases the width of the TMD of almost one order
of magnitude and leads to an $x$ dependence of the width that is different
from the original one, with no strong difference between the two versions
of the model.

\section{Conclusions}
\label{conclusions}
We have performed a study of the pion using light-front holographic QCD, 
which allows us to construct pion LFWFs. 
We took into consideration two different versions of the pion
LFWFs: pure-valence and effective-valence. 
For each version, 
the model contains three free parameters: the mass parameter $\kappa$
(expressing the strength of the confining harmonic
potential that breaks conformal invariance), the quark mass, and the scale of
the model. 
We fix the parameters by
comparison to experimental
information on pion form factors and PDFs. 

We obtain a value of $\kappa$ in
agreement with previous estimate~\citep{Brodsky:2007hb}, 
for the pure-valence version of the model. 
For the effective-valence version, we obtain a
smaller value~\citep{Gutsche:2013zia,Gutsche:2014zua}. The best agreement with data in the case of massive quarks is obtained for a quark mass
$m=0.05$ GeV for the pure-valence version, and $m=0.005$ GeV for the effective-valence
version. 
In order to achieve a fair agreement with the pion PDF at 5 GeV, 
the model scale has to be set to about $0.5$ GeV. 
This turns out to be true both for the pure-valence and the effective-valence
LFWF.

The sets of parameters obtained have then been used to study the unpolarized 
TMD of the pion. At the model scale, the resulting TMD has a Gaussian shape
with a width (defined as the position of the peak of the distributions
$k_{\bot}f_1(x,k^2_{\bot})$) of about 0.1 GeV at $x=0.5$. The $x$ dependence of
this width is different in the two versions of the model: in the pure-valence
model the TMD attains its maximal width at $x=0.5$; in the effective model,
this happens at $x=0.28$. After the TMD is
evolved to a typical experimental scale of about 5 GeV, its width increases 
by almost one order
of magnitude. The $x$ dependence is different from the one at the model scale:
the width
grows monotonically as $x$ decreases, and 
the differences between the two versions
of the model fade away.

\section*{Acknowledgements}
The authors are grateful to S.J. Brodsky, G.F. de Teramond and P.J.G. Mulders for stimulating discussions.
This work was supported by the European Research
Council (ERC) under the programs QWORK (contract No. 320389) and 3DSPIN (contract No. 647981).


\end{document}